%% file: main.tex
\documentclass[a4paper,11pt]{article}
\usepackage{pos}
\input{packages}

\title{Scale setting for $\mathcal{N} = 1$ SUSY Yang-Mills at large-$N$ through volume-reduced twisted matrix model}

\author*[a,b]{P. Butti}
\author[b]{M. Garcia Perez}
\author[a,b]{A. Gonzalez-Arroyo}
\author[c,d]{K. I. Ishikawa}
\author[d]{M. Okawa}

\affiliation[a]{Departamento de Fisica Teorica, Modulo 15, Universidad Autonoma de Madrid,
Cantoblanco, E-28045, Madrid, Spain}

\affiliation[b]{Instituto de Fisica Teorica UAM-CSIC,
  Calle Nicolas Cabrera 13-15, Universidad Autonoma de Madrid, Cantoblanco, E-28049, Madrid, Spain}
 
\affiliation[c]{Core of Research for the Energetic Universe, Graduate School of Advanced Science and Engineering, Hiroshima University, Higashi-Hiroshima, Hiroshima 739-8526, Japan}

\affiliation[d]{Graduate School of Advanced Science and Engineering, Hiroshima University, Higashi-Hiroshima, Hiroshima 739-8526, Japan}

\emailAdd{pietro.butti@uam.es}

\abstract{$\mathcal{N}=1$ SUSY Yang-Mills theory is an appealing theoretical framework that has been studied in the literature using different methods, including standard lattice simulations. Among these, the volume-reduced twisted Eguchi-Kawai model, endowed with one adjoint Majorana fermion, could play an important role in studying its large-$N$ limit via the Curci-Veneziano prescription. In this talk, we present our results on the analysis of the scale of the theory, performed via different methods based on purely gluonic observables as well as (quenched) fundamental mesons in the chiral limit. These lattice results will be used as a scale setting for the analysis of the spectrum of the theory.}

\FullConference{%
 The 38th International Symposium on Lattice Field Theory, LATTICE2021
  26th-30th July, 2021
  Zoom/Gather@Massachusetts Institute of Technology
}

\begin{document}
	
	\begin{flushright}
		\vspace*{-8em}
		FTUAM-21-4, HUPD-2109, IFT-UAM/CSIC-21-120
		\vspace*{4em}
	\end{flushright}

\maketitle

\input{tek_introduction}
\input{tek_flow}
\input{tek_meson}

\input{tek_conclusion}

\acknowledgments
P.B. M.G.P and A.G.A. acknowledge financial support from the MINECO/FEDER grant PGC2018-094857-B-I00 and the MINECO Centro de Excelencia Severo Ochoa Program SEV-2016-0597. This publication is supported by the European Project H2020-MSCAITN-2018-813942 (EuroPLEx) and the EU Horizon 2020 research and innovation programme STRONG-2020 project, under grant agreement No. 824093. 

M. O. is supported by JSPS KAKENHI Grant Numbers 21K03576 and 17K05417. K.I. I. is supported by MEXT as “Program for Promoting Researches on the Supercomputer Fugaku” (Simulation for basic science: from fundamental laws of particles to creation of nuclei, JPMXP1020200105) and JICFuS. 

The calculation have been done on Hydra cluster at IFT in Madrid and on the “Centro de Supercomputación de Galicia (CESGA)”. This work is based also on calculations performed on SX-ACE (Osaka U.) and Oakbridge-CX (U. of Tokyo) through the HPCI System Research Project
(Project ID: hp210027, hp200027, hp190004).

\end{document}

%% file: packages.tex
\usepackage[T1]{fontenc}    
\usepackage[utf8]{inputenc} 
\usepackage[english]{babel} 
\usepackage{lscape}         

\usepackage{amsmath}        
\usepackage{amsthm}         
\usepackage{amsfonts}		
\usepackage{amscd}          
\usepackage{braket}         
\usepackage{empheq}         
\usepackage{bm}             
\usepackage{physics}
\usepackage{dsfont}         
\usepackage{amsopn}         
\usepackage{mathrsfs}       
\usepackage{xfrac}          
\usepackage{bbold}          
\usepackage{slashed}        
\usepackage{mathtools}      

\usepackage{booktabs}       
\usepackage{multirow}       
\usepackage{multicol}

\usepackage{graphicx}
\usepackage{pgfplots}
    \pgfplotsset{compat=1.13}
\usepackage{tikz}
\usepackage{subfigure}
\usepackage{import}
\usepackage{calc}
\usepackage[font=small,format=hang]{caption} 

\usepackage[babel]{csquotes}

\usepackage{footmisc}           
\usepackage{cancel}             
\usepackage[nowrite,infront,swapnames]{frontespizio} 
\usepackage[version=4]{mhchem}  
\usepackage{xcolor}
    \definecolor{Goldenrod}{rgb}{0.85, 0.65, 0.13}
    \definecolor{NavyBlue}{rgb}{0.0, 0.0, 0.5}
    \definecolor{BrickRed}{rgb}{0.8, 0.0, 0.0}
    \definecolor{OliveGreen}{rgb}{0.33, 0.42, 0.18}
\usepackage{algorithm}          
\usepackage[all,cmtip]{xy}      
\usepackage{feynmf}
\usepackage{todonotes}    

\usepackage{listings}           
    \lstnewenvironment{cppcode}{\lstset{frame=single,language=C++,basicstyle=\small\ttfamily,keywordstyle=\color{blue},commentstyle=\color{darkspringgreen},stringstyle=\color{magenta},backgroundcolor=\color{mygray}}}{}
\usepackage{fancyhdr}
    \fancypagestyle{plain}{
    	\fancyhf{}
    	\rfoot{\thepage}

    }

%% file: tek_introduction.tex
\section{Introduction}
$\mathcal{N}=1$ SUSY Yang-Mills theory is the simplest supersymmetric extension of the gluonic sector of the Standard Model, whose dynamical degrees of freedom are gluons and gluinos, which are Majorana fermions in the adjoint representation. The most interesting regime of the theory is non-perturbative, and a lot of interesting phenomena are expected to occur: formation of super-multiplets of bound states, confinement and a non-trivial vacuum structure. The $SU(3)$ version of the theory is interesting for extensions of the Standard Model and has been subject of interest of numerous investigations also on the lattice (see \cite{progressSUSY,Ali:2019agk} and references therein). Still, the purpose of this work is to study the large-$N$ limit of this theory.
Adjoint fermions like gluinos must be treated dynamically in lattice simulations, resulting in a too heavy computational effort using standard lattice techniques.

Our method consists in exploiting volume reduction at large-$N$, using twisted space-time reduced models on a $1^4$ lattice with $N=169,289,361$ at three different values of the lattice spacing \cite{TEKadj2}. In this framework, the finite-$N$ corrections of the twisted model amount to finite volume corrections of an ordinary $(L=\sqrt{N})^4$ lattice gauge theory. The resulting action is
\begin{equation}
    S = -bN\sum_{\mu\neq\nu} \text{tr}[z_{\mu\nu} U_\mu U_\nu U^\dagger_\mu U^\dagger_\mu]- \frac{1}{2}\bar{\psi}D_w\psi
\end{equation}
where $U_\mu$ are four $SU(N)$ link variables, $b$ is the inverse of the lattice 't Hooft coupling $b=\sfrac{1}{g^2 N}=\sfrac{1}{\lambda}$, $\psi$ are Grassman spinors transforming in the adjoint representation and $D_w$ is the Wilson-Dirac matrix for Wilson fermions, while $z_{\mu\nu}=e^{i\frac{2\pi k}{\sqrt{N}}}$ when $\mu<\nu$ for $k$ and $\sqrt{N}$ coprime. Using RHMC algorithm adapted to our case, we were able to generate ensembles of configurations for several values of the 't Hooft coupling and several gluino masses. A more detailed discussion about the methodology was presented at this conference by K. I. Ishikawa and can be found in \cite{pos}, in which the generation of configurations is tackled and some preliminary results about the adjoint-spectrum are presented. 

In this work we will perform the scale setting for this theory and explore the supersymmetric limit of vanishing gluino mass. We will use standard Wilson flow techniques adapted to the twisted framework, cross-checking them with other natural scales of the theory.

\subsection{Some comments on the SUSY-$\chi$ limit}\label{mpcac}
It is well-known that lattice discretizations explicitly break supersymmetry, as well as a non vanishing gluino mass term in the action. Nevertheless, as shown by Curci and Veneziano \cite{CurciVeneziano}, supersymmetry and chiral symmetry are restored in the continuum limit by the same tuning of the bare gluino mass. Hence, in order to study SUSY on the lattice, one has to extrapolate all results to the limit of vanishing renormalized quark mass $m_q^{(r)}$.
We briefly remind that in QCD, spontaneous breaking of chiral symmetry by the fermion condensate ensures that the chiral limit can be achieved by tuning $\pi$-meson mass to 0, since Gell-Mann-Oakes-Renner relation ensures that $m_\pi^2\propto m_q^{(r)}$. Pions are indeed standard (pseudo-)Goldstone boson for this mechanism, therefore can be used to achieve the chiral limit. On the other hand, for $\mathcal{N}=1$ SUSY Yang-Mills, the corresponding chiral symmetry group is already anomalously broken and only a remnant \textit{discrete} subgroup breaks spontaneously in the vacuum., i.e. no Goldstone bosons are present in the spectrum. 

Nevertheless, one method to achieve the chiral limit \cite{Ali:2019agk} employs the mass of the adjoint-$\pi$, defined as the connected part of the adj-$\eta'$ correlator. Although not being present in the physical spectrum of the theory, the mass $m_{\text{adj}-\pi}^2$ is expected to vanish with the renormalized gluino mass at the chiral limit\footnote{This is suggested by arguments based on partially-quenched chiral perturbation theory and OZI-approximation considerations.}. Analogously, one can define a PCAC mass $am_\text{PCAC}$ with the usual definition involving pseudo-scalar and an axial currents and then tuning it to 0. We will use this last method to approach the chiral limit.

%% file: tek_flow.tex
\section{Scales from twisted Wilson flow}
Using standard methodology, one can consider the evolution of the gauge field configuration driven by the Wilson flow equation ($\partial_t A_\mu(x;t) = D_\nu G_{\mu\nu}(x;t) $). A common observable is the \textit{flowed energy density} $E(t)$
\begin{equation}\label{energydensity}
    \expval{E(t)} = \frac{1}{2}\expval{\Tr G_{\mu\nu}(x,t)G_{\mu\nu}(x,t)}
\end{equation}
$t$ being the flow time. In our one-site reduced lattice different equivalent discretized versions of \eqref{energydensity} are possible. We adopt the clover version of the field strength
\begin{equation}
    \hat{E} = -\frac{1}{128}   \sum_{\mu,\nu}
    \Tr[  z_{\nu\mu} U_\nu U_\mu U^\dagger_\nu U^\dagger_\mu  
    + z_{\nu\mu} U_\mu U^\dagger_\nu U^\dagger_\mu U_\nu
    + z_{\nu\mu}  U^\dagger_\nu U^\dagger_\mu U_\nu U_\mu
    + z_{\nu\mu}  U^\dagger_\mu U_\nu U_\mu U^\dagger_\mu  
    - \text{h.c.}]^2
\end{equation}
to evaluate on the lattice the dimensionless flowed energy density $\Phi(T,N)=\sfrac{1}{N}\expval{T^2 \hat{E}(T)}$ where $T=\frac{t}{a^2}$. Several reference scales can be found by solving the following implicit equations 
\begin{align}\label{t0}
    \eval{\frac{1}{N}\expval{T^2 \hat{E}(T)}}_{T=T_0 (T_1)} &= 0.1 (0.05) \\\label{w0}
    \eval{T\dv{T}\qty(\frac{1}{N}\expval{T^2 \hat{E}(T)})}_{T=W_0^2 (W_1^2)} &= 0.1 (0.05)
\end{align} 
where $T_i=\sfrac{t_i}{a^2}$, $W_i=\sfrac{w_i}{a}$. We remind that, due to the extra $\sfrac{1}{N}$ factor in front of the flowed energy density, $T_0$ (as well as $W_0$) corresponds to the usual definition given in standard $SU(3)$ lattice QCD simulations. On the other hand $T_1$ and $W_1$ are chosen with an arbitrary reference scale 0.05.

Finite-$N$ effects are a big source of systematic errors and they have to be taken under control. On the reduced lattice, they are deeply related with finite-size effect on a ordinary lattice. In fact, Wilson flow smears gauge variables on a volume $(\sqrt{8T})^4$, which could become comparable with the volume of the (effective) lattice as the flow time is increased.  These effects are already present in perturbation theory and have been quantified up to $\order{\lambda^2}$ \cite{flow}. At leading order, finite-$N$ contributions are accounted for by an overall coefficient $\hat{\mathcal{N}}_N(\sqrt{\sfrac{8T}{N}})$ that can be evaluated analytically. Our idea is to substitute this coefficient with its continuum infinite-volume version $\hat{\mathcal{N}}_\infty(0)=\frac{3}{128\pi^2}$.
In order to do that, we define a new flowed energy density as
\begin{equation}
    \Phi'(T) = \frac{3}{128\pi^2\hat{\mathcal{N}}_N(T)}\Phi(T,N)
\end{equation}
where an exact analytical expression for $\hat{\mathcal{N}}_N(\sfrac{\sqrt{8T}}{\sqrt{N}})$ can be found in \cite{flow}. This overall factor should remove both finite-$N$ contributions and lattice artifact effects to the flowed energy density calculated at tree-level in perturbation theory. Once corrected we fit each curve to a parabola in the flow-time range $[1.5,\frac{N}{128}]$\footnote{The lower bounds is to get rid of lattice artifacts, while the upper corresponds to $\sqrt{8T}<\sfrac{\sqrt{N}}{4}$, where the smearing radius is reasonably smaller than the total volume of the effective lattice.} 

\begin{figure}
    \centering
    \scalebox{0.75}{\input{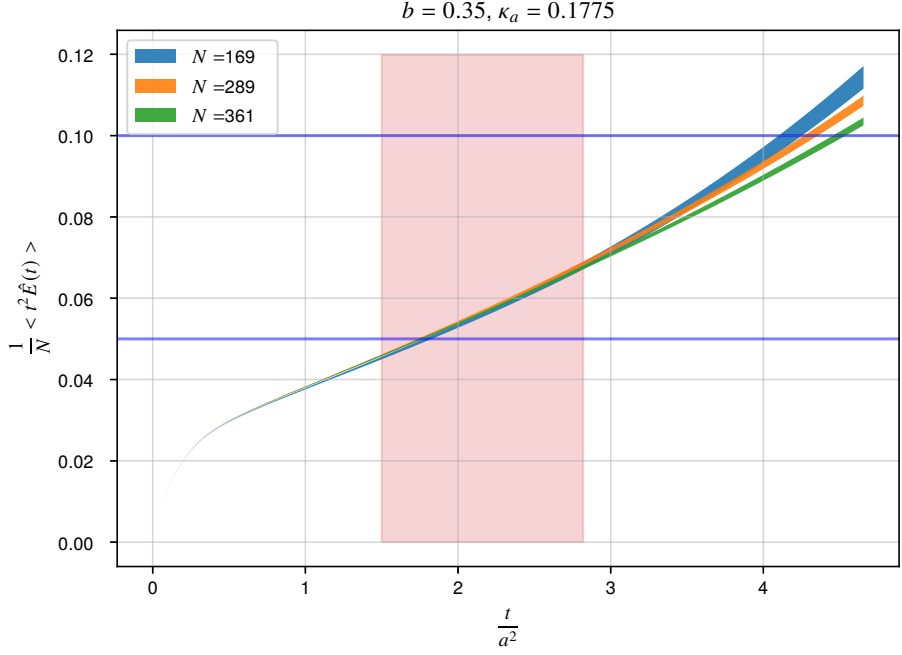}}
    \caption{Examples of the flowed energy density after norm correction as a function of $\frac{t}{a^2}$. The blue horizontal lines represent the values $0.05$ and $0.1$ at which the curves are cut to extract $T_0$ and $T_1$. The vertical red band represents the biggest fitting region for the $N=361$ curve, whose upper limit is $~2.82$. Smaller values of $N$ has smaller upper limits.   
    As can be seen, within corresponding fitting region, the corrected curves are independent on the number of colors.}
    \label{fig:renorming}
\end{figure}

This method results to be very effective inside a properly defined fitting window, as can be seen from Fig. \ref{fig:renorming}. For this choice of parameters $\sfrac{t_1}{a^2}$ falls inside the fitting region, while $\sfrac{t_0}{a^2}$ can only be obtained through an extrapolation. In these cases, nicely exemplified by the plot, the corrected flows at different values of $N$ collapse to a unique curve, as the dependence on $N$ has been removed. On the other hand, outside the fitting region for large flow-times, a remnant $N$-dependence manifests. As a matter of fact, when approaching the critical region of the bare mass parameter ($\kappa_c\gtrsim\kappa_a$), the extraction of the masses results to be more complicated.

However, among the four different scales we calculated, it turns out that $\sqrt{8t_1}$ is far less affected by finite-$N$ effect and has the smallest uncertainties. Furthermore, we observed that the dimensionless ratios $R=\sfrac{\sqrt{8t_0}}{\sqrt{8t_1}}$, $R_0=\sfrac{w_0}{\sqrt{8t_1}}$ and $R=\sfrac{w_1}{\sqrt{8t_1}}$ are constant in a region where $\sfrac{1}{\sqrt{8T_1}}<0.3$ and $\sfrac{\sqrt{N}}{\sqrt{8T_1}}>3.5$, as one would expect from scaling. For this reason, we fix those ratios to the average value they assume in this region and use them to convert our results to the other units. For example:
\begin{equation}
    \sqrt{8T_0(b,\kappa_a)}= R\sqrt{8T_1(b,\kappa_a)}
\end{equation}
(analogous relations hold for $W_i$). The values of these ratios are $R=1.602(20)$, $R_0=\sfrac{w_0}{\sqrt{8t_1}}=0.567(12)$, $R_1=\sfrac{w_1}{\sqrt{8t_1}}=0.4511(75)$. 


\subsection{Results and some comments}
\begin{figure}
    \centering
    \input{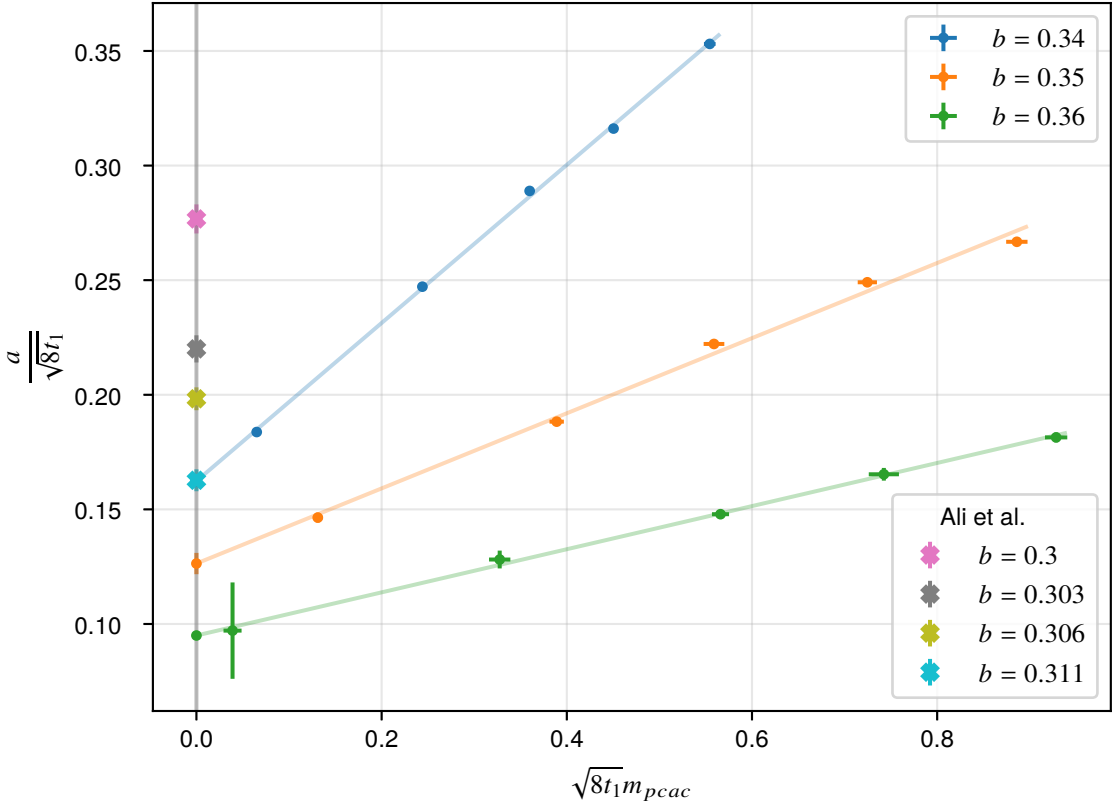}
    \caption{Lattice spacing as a function of $m_\text{PCAC}$ in units of $\sqrt{8t_1}$. Each color represents a different gauge coupling $b$, for both the points and the fitted curves. The extrapolated values lying on the grey axes correspond to the chiral limit where $\kappa_a=\kappa_c$. 
    On the grey line, we also report the extrapolated chiral value of $\sqrt{8t_1} = \sfrac{w_0}{R_0}$, where $w_0$ were taken from \cite{Ali:2019agk}.}
    \label{fig:aw0}
\end{figure}

In Fig. \ref{fig:aw0} we plot the lattice spacing in units of $\sqrt{8t_1}$ as a function of the $m_\text{PCAC}$ mass in the same units.
As shown in the picture, we fitted for each value of $b$ our points to a straight line and extrapolated the value of the lattice spacing in units of $\sqrt{8t_1}$ to the chiral limit $m_\text{PCAC}\rightarrow 0$. Remarkably, we notice that a linear extrapolation is good enough, as signalled by low $\chi^2$ and no higher order polynomial are needed. We report in Tab. \ref{aw0} our preliminary values of the lattice spacing in units of $\sqrt{8t_1}$ extrapolated to the chiral limit.
\begin{table}[h!]
    \centering
    \begin{tabular}{clc}
    \toprule
        $b$ & SUSY $\frac{a}{\sqrt{8t_1}}$ & $\frac{a}{w_0}$ \\
    \midrule
        0.34 & 0.1634(23) & 0.2882(40)\\
        0.35 & 0.1283(65) & 0.226(11)\\
        0.36 & 0.0972(44) & 0.1714(77)\\
    \bottomrule
    \end{tabular}
    \caption{Preliminary values of the lattice spacings for different gauge couplings $b$ extrapolated to the chiral limit, where SUSY is expected to be restored. Errors are given qualitatively by assigning to the extrapolated value the biggest uncertainty among the fitting points. For the sake of completeness, in the third column we report the lattice spacing in $w_0$ units, dividing the second column values by $R_0$.}
    \label{aw0}
\end{table}


Having computed the scale from our data, it is interesting to compare it with the theoretical expectations. The dependence of the scale on the coupling is dictated by the $\beta$-function. This can be computed in perturbation theory to two loops and the result for $SU(N)$ gauge theory with $N_f$ flavours of adjoint Dirac quarks is:
\begin{equation}\label{beta}
    a\dv{\lambda(a)}{a} = - \beta(\lambda) = b_0 \lambda^2 + b_1 \lambda^3 + \order{\lambda^4}\quad\text{ with } b_0 = -\frac{4N_f-11}{24\pi^2}\,\text{ and } b_1 = -\frac{16N_f-17}{192\pi^4}
\end{equation}
In our case we have one Majorana fermion corresponding to $N_f=\sfrac{1}{2}$. After integrating the equation we get a prediction for the dependence of the logarithm of the scale on the coupling, up to an additive constant giving the corresponding lambda parameter $\Lambda$. However, it is well-known that, on the lattice, perturbative scaling works better using an improved lattice coupling constant $\lambda_I=1/b_I(b)$. Many
options  have been put forward and tested in the literature. For definiteness we will stick to  the one proposed by Allton-Teper-Trivini \cite{allton}: $b_I=bP(b)$, where $P(b)$ is the average value of the plaquette (other choices have been tested and give similar results). With this choice our scale results match nicely with the predicted behaviour with a fitted  Lambda parameter (The square root of the chi-square per degree of freedom being 0.68). Alternatively, we can fit the two loop formula but leaving also the first
coefficient of the beta function as a free parameter. Our best fit gives $b_0=0.0408(33)$ which is perfectly compatible with the exact result $b_0=\sfrac{3}{8 \pi^2}=0.0380$.


%% file: tek_meson.tex
\section{Scales from fundamental quenched mesons}
Although adjoint-quarks have to be treated dynamically, we recall that at large-$N$ fundamental quark loops are naturally suppressed and the quenched approximation is exact. These fundamental quarks ($\kappa_\text{fund}$) form bound states, e.g. $\pi$ or $\rho$-like states, and do not influence the dynamics. The mass of these mesons depends on an additional scale, i.e. the fundamental quarks mass, but as we take the fundamental chiral limit i.e. $\kappa_\text{fund}\rightarrow\kappa_\text{fund}^{(c)}$ the resulting $\rho$-meson mass can be considered a natural scale which can be compared with $\sqrt{8t_1}$.


\subsection{Spectroscopy of fundamental light-meson sector}
We refer to \cite{mesonN} (in particular sections 3.2 and 3.3) and references therein for details of the procedure we followed. We calculated correlators for several fundamental quark masses and for different $(b,\kappa_a)$ following the strategy in the reference quoted, and creating several operators with the desired quantum numbers by applying different level of combined APE+Wuppertal smearing, then we solved the related GEVP problem to disentangle the signal of the fundamental state.
This procedure allowed us to extract masses  $am_\pi$, $am_\rho$ and $am_\text{pcac}^\text{fund}$ from the hyperbolic cosine fall-offs of the resulting correlator. As shown in Fig. \ref{fig:mrho_mpcac}, chiral $\rho$-meson masses $am_\rho^\chi$ are extracted tuning $am_\text{pcac}^\text{fund}\rightarrow0$. The errors on the extrapolated values have been calculated using a standard jackknife procedure, and result to be of almost 10\%, which is not as precise as what we obtained for the Wilson flow related scales. Nevertheless our main goal here is to check the consistency of our method with several scales coming from different frameworks. This is achieved in Fig. \ref{fig:mrhost1}, where we plotted the dimensionless quantity $m_\rho^\chi\sqrt{8t_1}$ for some of the cases we analyzed. As expected, the behavior is constant within the errors. 
\begin{figure}
    \scalebox{0.85}{\input{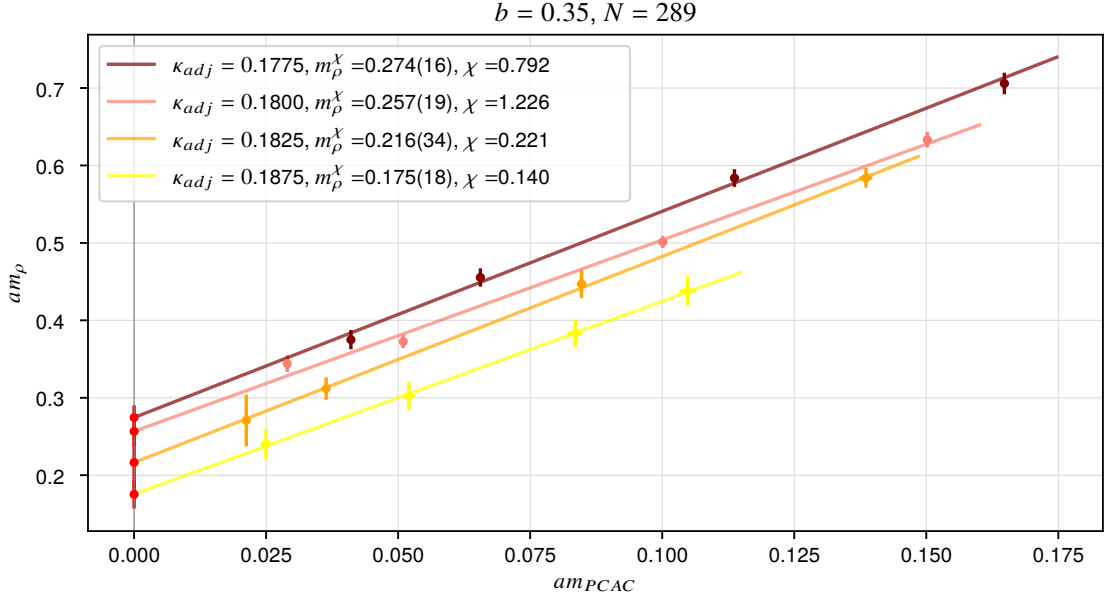}}
    \caption{Extrapolation of the $\rho$-meson masses to the chiral limit $am_\text{pcac}\rightarrow0$ for $b=0.35$. Errors on the extrapolated value are calculated with standard jackknife procedure.}
    \label{fig:mrho_mpcac}
\end{figure}
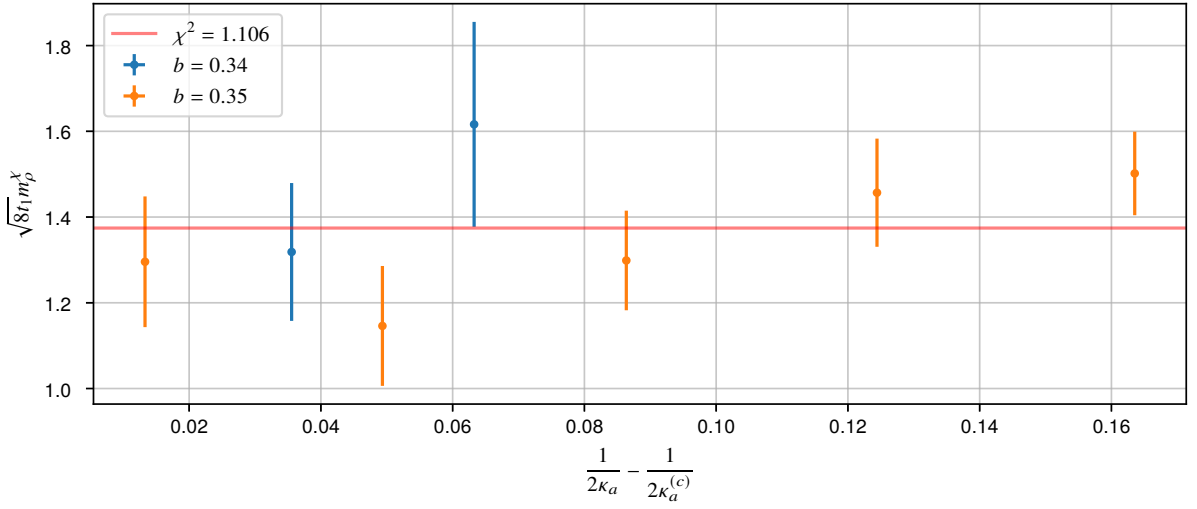
\begin{figure}
    \centering
    \scalebox{0.8}{\input{PLOT/mrhost1.pgf}}
    \caption{Dimensionless product $m_\rho^\chi\sqrt{8t_1}$ for different value of $b$ and for the adjoint meson mass. The red line is the average of the plotted points.}
    \label{fig:mrhost1}
\end{figure}

%% file: PLOT/mrhost1.pgf
\begingroup%
\makeatletter%
\begin{pgfpicture}%
\pgfpathrectangle{\pgfpointorigin}{\pgfqpoint{8.000000in}{3.500000in}}%
\pgfusepath{use as bounding box, clip}%
\begin{pgfscope}%
\pgfsetbuttcap%
\pgfsetmiterjoin%
\definecolor{currentfill}{rgb}{1.000000,1.000000,1.000000}%
\pgfsetfillcolor{currentfill}%
\pgfsetlinewidth{0.000000pt}%
\definecolor{currentstroke}{rgb}{1.000000,1.000000,1.000000}%
\pgfsetstrokecolor{currentstroke}%
\pgfsetdash{}{0pt}%
\pgfpathmoveto{\pgfqpoint{0.000000in}{0.000000in}}%
\pgfpathlineto{\pgfqpoint{8.000000in}{0.000000in}}%
\pgfpathlineto{\pgfqpoint{8.000000in}{3.500000in}}%
\pgfpathlineto{\pgfqpoint{0.000000in}{3.500000in}}%
\pgfpathclose%
\pgfusepath{fill}%
\end{pgfscope}%
\begin{pgfscope}%
\pgfsetbuttcap%
\pgfsetmiterjoin%
\definecolor{currentfill}{rgb}{1.000000,1.000000,1.000000}%
\pgfsetfillcolor{currentfill}%
\pgfsetlinewidth{0.000000pt}%
\definecolor{currentstroke}{rgb}{0.000000,0.000000,0.000000}%
\pgfsetstrokecolor{currentstroke}%
\pgfsetstrokeopacity{0.000000}%
\pgfsetdash{}{0pt}%
\pgfpathmoveto{\pgfqpoint{0.777361in}{0.737281in}}%
\pgfpathlineto{\pgfqpoint{7.850000in}{0.737281in}}%
\pgfpathlineto{\pgfqpoint{7.850000in}{3.350000in}}%
\pgfpathlineto{\pgfqpoint{0.777361in}{3.350000in}}%
\pgfpathclose%
\pgfusepath{fill}%
\end{pgfscope}%
\begin{pgfscope}%
\pgfpathrectangle{\pgfqpoint{0.777361in}{0.737281in}}{\pgfqpoint{7.072639in}{2.612719in}}%
\pgfusepath{clip}%
\pgfsetrectcap%
\pgfsetroundjoin%
\pgfsetlinewidth{0.803000pt}%
\definecolor{currentstroke}{rgb}{0.690196,0.690196,0.690196}%
\pgfsetstrokecolor{currentstroke}%
\pgfsetstrokeopacity{0.700000}%
\pgfsetdash{}{0pt}%
\pgfpathmoveto{\pgfqpoint{1.396431in}{0.737281in}}%
\pgfpathlineto{\pgfqpoint{1.396431in}{3.350000in}}%
\pgfusepath{stroke}%
\end{pgfscope}%
\begin{pgfscope}%
\pgfsetbuttcap%
\pgfsetroundjoin%
\definecolor{currentfill}{rgb}{0.000000,0.000000,0.000000}%
\pgfsetfillcolor{currentfill}%
\pgfsetlinewidth{0.803000pt}%
\definecolor{currentstroke}{rgb}{0.000000,0.000000,0.000000}%
\pgfsetstrokecolor{currentstroke}%
\pgfsetdash{}{0pt}%
\pgfsys@defobject{currentmarker}{\pgfqpoint{0.000000in}{-0.048611in}}{\pgfqpoint{0.000000in}{0.000000in}}{%
\pgfpathmoveto{\pgfqpoint{0.000000in}{0.000000in}}%
\pgfpathlineto{\pgfqpoint{0.000000in}{-0.048611in}}%
\pgfusepath{stroke,fill}%
}%
\begin{pgfscope}%
\pgfsys@transformshift{1.396431in}{0.737281in}%
\pgfsys@useobject{currentmarker}{}%
\end{pgfscope}%
\end{pgfscope}%
\begin{pgfscope}%
\definecolor{textcolor}{rgb}{0.000000,0.000000,0.000000}%
\pgfsetstrokecolor{textcolor}%
\pgfsetfillcolor{textcolor}%
\pgftext[x=1.396431in,y=0.640058in,,top]{\color{textcolor}\sffamily\fontsize{10.000000}{12.000000}\selectfont 0.02}%
\end{pgfscope}%
\begin{pgfscope}%
\pgfpathrectangle{\pgfqpoint{0.777361in}{0.737281in}}{\pgfqpoint{7.072639in}{2.612719in}}%
\pgfusepath{clip}%
\pgfsetrectcap%
\pgfsetroundjoin%
\pgfsetlinewidth{0.803000pt}%
\definecolor{currentstroke}{rgb}{0.690196,0.690196,0.690196}%
\pgfsetstrokecolor{currentstroke}%
\pgfsetstrokeopacity{0.700000}%
\pgfsetdash{}{0pt}%
\pgfpathmoveto{\pgfqpoint{2.249184in}{0.737281in}}%
\pgfpathlineto{\pgfqpoint{2.249184in}{3.350000in}}%
\pgfusepath{stroke}%
\end{pgfscope}%
\begin{pgfscope}%
\pgfsetbuttcap%
\pgfsetroundjoin%
\definecolor{currentfill}{rgb}{0.000000,0.000000,0.000000}%
\pgfsetfillcolor{currentfill}%
\pgfsetlinewidth{0.803000pt}%
\definecolor{currentstroke}{rgb}{0.000000,0.000000,0.000000}%
\pgfsetstrokecolor{currentstroke}%
\pgfsetdash{}{0pt}%
\pgfsys@defobject{currentmarker}{\pgfqpoint{0.000000in}{-0.048611in}}{\pgfqpoint{0.000000in}{0.000000in}}{%
\pgfpathmoveto{\pgfqpoint{0.000000in}{0.000000in}}%
\pgfpathlineto{\pgfqpoint{0.000000in}{-0.048611in}}%
\pgfusepath{stroke,fill}%
}%
\begin{pgfscope}%
\pgfsys@transformshift{2.249184in}{0.737281in}%
\pgfsys@useobject{currentmarker}{}%
\end{pgfscope}%
\end{pgfscope}%
\begin{pgfscope}%
\definecolor{textcolor}{rgb}{0.000000,0.000000,0.000000}%
\pgfsetstrokecolor{textcolor}%
\pgfsetfillcolor{textcolor}%
\pgftext[x=2.249184in,y=0.640058in,,top]{\color{textcolor}\sffamily\fontsize{10.000000}{12.000000}\selectfont 0.04}%
\end{pgfscope}%
\begin{pgfscope}%
\pgfpathrectangle{\pgfqpoint{0.777361in}{0.737281in}}{\pgfqpoint{7.072639in}{2.612719in}}%
\pgfusepath{clip}%
\pgfsetrectcap%
\pgfsetroundjoin%
\pgfsetlinewidth{0.803000pt}%
\definecolor{currentstroke}{rgb}{0.690196,0.690196,0.690196}%
\pgfsetstrokecolor{currentstroke}%
\pgfsetstrokeopacity{0.700000}%
\pgfsetdash{}{0pt}%
\pgfpathmoveto{\pgfqpoint{3.101937in}{0.737281in}}%
\pgfpathlineto{\pgfqpoint{3.101937in}{3.350000in}}%
\pgfusepath{stroke}%
\end{pgfscope}%
\begin{pgfscope}%
\pgfsetbuttcap%
\pgfsetroundjoin%
\definecolor{currentfill}{rgb}{0.000000,0.000000,0.000000}%
\pgfsetfillcolor{currentfill}%
\pgfsetlinewidth{0.803000pt}%
\definecolor{currentstroke}{rgb}{0.000000,0.000000,0.000000}%
\pgfsetstrokecolor{currentstroke}%
\pgfsetdash{}{0pt}%
\pgfsys@defobject{currentmarker}{\pgfqpoint{0.000000in}{-0.048611in}}{\pgfqpoint{0.000000in}{0.000000in}}{%
\pgfpathmoveto{\pgfqpoint{0.000000in}{0.000000in}}%
\pgfpathlineto{\pgfqpoint{0.000000in}{-0.048611in}}%
\pgfusepath{stroke,fill}%
}%
\begin{pgfscope}%
\pgfsys@transformshift{3.101937in}{0.737281in}%
\pgfsys@useobject{currentmarker}{}%
\end{pgfscope}%
\end{pgfscope}%
\begin{pgfscope}%
\definecolor{textcolor}{rgb}{0.000000,0.000000,0.000000}%
\pgfsetstrokecolor{textcolor}%
\pgfsetfillcolor{textcolor}%
\pgftext[x=3.101937in,y=0.640058in,,top]{\color{textcolor}\sffamily\fontsize{10.000000}{12.000000}\selectfont 0.06}%
\end{pgfscope}%
\begin{pgfscope}%
\pgfpathrectangle{\pgfqpoint{0.777361in}{0.737281in}}{\pgfqpoint{7.072639in}{2.612719in}}%
\pgfusepath{clip}%
\pgfsetrectcap%
\pgfsetroundjoin%
\pgfsetlinewidth{0.803000pt}%
\definecolor{currentstroke}{rgb}{0.690196,0.690196,0.690196}%
\pgfsetstrokecolor{currentstroke}%
\pgfsetstrokeopacity{0.700000}%
\pgfsetdash{}{0pt}%
\pgfpathmoveto{\pgfqpoint{3.954690in}{0.737281in}}%
\pgfpathlineto{\pgfqpoint{3.954690in}{3.350000in}}%
\pgfusepath{stroke}%
\end{pgfscope}%
\begin{pgfscope}%
\pgfsetbuttcap%
\pgfsetroundjoin%
\definecolor{currentfill}{rgb}{0.000000,0.000000,0.000000}%
\pgfsetfillcolor{currentfill}%
\pgfsetlinewidth{0.803000pt}%
\definecolor{currentstroke}{rgb}{0.000000,0.000000,0.000000}%
\pgfsetstrokecolor{currentstroke}%
\pgfsetdash{}{0pt}%
\pgfsys@defobject{currentmarker}{\pgfqpoint{0.000000in}{-0.048611in}}{\pgfqpoint{0.000000in}{0.000000in}}{%
\pgfpathmoveto{\pgfqpoint{0.000000in}{0.000000in}}%
\pgfpathlineto{\pgfqpoint{0.000000in}{-0.048611in}}%
\pgfusepath{stroke,fill}%
}%
\begin{pgfscope}%
\pgfsys@transformshift{3.954690in}{0.737281in}%
\pgfsys@useobject{currentmarker}{}%
\end{pgfscope}%
\end{pgfscope}%
\begin{pgfscope}%
\definecolor{textcolor}{rgb}{0.000000,0.000000,0.000000}%
\pgfsetstrokecolor{textcolor}%
\pgfsetfillcolor{textcolor}%
\pgftext[x=3.954690in,y=0.640058in,,top]{\color{textcolor}\sffamily\fontsize{10.000000}{12.000000}\selectfont 0.08}%
\end{pgfscope}%
\begin{pgfscope}%
\pgfpathrectangle{\pgfqpoint{0.777361in}{0.737281in}}{\pgfqpoint{7.072639in}{2.612719in}}%
\pgfusepath{clip}%
\pgfsetrectcap%
\pgfsetroundjoin%
\pgfsetlinewidth{0.803000pt}%
\definecolor{currentstroke}{rgb}{0.690196,0.690196,0.690196}%
\pgfsetstrokecolor{currentstroke}%
\pgfsetstrokeopacity{0.700000}%
\pgfsetdash{}{0pt}%
\pgfpathmoveto{\pgfqpoint{4.807443in}{0.737281in}}%
\pgfpathlineto{\pgfqpoint{4.807443in}{3.350000in}}%
\pgfusepath{stroke}%
\end{pgfscope}%
\begin{pgfscope}%
\pgfsetbuttcap%
\pgfsetroundjoin%
\definecolor{currentfill}{rgb}{0.000000,0.000000,0.000000}%
\pgfsetfillcolor{currentfill}%
\pgfsetlinewidth{0.803000pt}%
\definecolor{currentstroke}{rgb}{0.000000,0.000000,0.000000}%
\pgfsetstrokecolor{currentstroke}%
\pgfsetdash{}{0pt}%
\pgfsys@defobject{currentmarker}{\pgfqpoint{0.000000in}{-0.048611in}}{\pgfqpoint{0.000000in}{0.000000in}}{%
\pgfpathmoveto{\pgfqpoint{0.000000in}{0.000000in}}%
\pgfpathlineto{\pgfqpoint{0.000000in}{-0.048611in}}%
\pgfusepath{stroke,fill}%
}%
\begin{pgfscope}%
\pgfsys@transformshift{4.807443in}{0.737281in}%
\pgfsys@useobject{currentmarker}{}%
\end{pgfscope}%
\end{pgfscope}%
\begin{pgfscope}%
\definecolor{textcolor}{rgb}{0.000000,0.000000,0.000000}%
\pgfsetstrokecolor{textcolor}%
\pgfsetfillcolor{textcolor}%
\pgftext[x=4.807443in,y=0.640058in,,top]{\color{textcolor}\sffamily\fontsize{10.000000}{12.000000}\selectfont 0.10}%
\end{pgfscope}%
\begin{pgfscope}%
\pgfpathrectangle{\pgfqpoint{0.777361in}{0.737281in}}{\pgfqpoint{7.072639in}{2.612719in}}%
\pgfusepath{clip}%
\pgfsetrectcap%
\pgfsetroundjoin%
\pgfsetlinewidth{0.803000pt}%
\definecolor{currentstroke}{rgb}{0.690196,0.690196,0.690196}%
\pgfsetstrokecolor{currentstroke}%
\pgfsetstrokeopacity{0.700000}%
\pgfsetdash{}{0pt}%
\pgfpathmoveto{\pgfqpoint{5.660196in}{0.737281in}}%
\pgfpathlineto{\pgfqpoint{5.660196in}{3.350000in}}%
\pgfusepath{stroke}%
\end{pgfscope}%
\begin{pgfscope}%
\pgfsetbuttcap%
\pgfsetroundjoin%
\definecolor{currentfill}{rgb}{0.000000,0.000000,0.000000}%
\pgfsetfillcolor{currentfill}%
\pgfsetlinewidth{0.803000pt}%
\definecolor{currentstroke}{rgb}{0.000000,0.000000,0.000000}%
\pgfsetstrokecolor{currentstroke}%
\pgfsetdash{}{0pt}%
\pgfsys@defobject{currentmarker}{\pgfqpoint{0.000000in}{-0.048611in}}{\pgfqpoint{0.000000in}{0.000000in}}{%
\pgfpathmoveto{\pgfqpoint{0.000000in}{0.000000in}}%
\pgfpathlineto{\pgfqpoint{0.000000in}{-0.048611in}}%
\pgfusepath{stroke,fill}%
}%
\begin{pgfscope}%
\pgfsys@transformshift{5.660196in}{0.737281in}%
\pgfsys@useobject{currentmarker}{}%
\end{pgfscope}%
\end{pgfscope}%
\begin{pgfscope}%
\definecolor{textcolor}{rgb}{0.000000,0.000000,0.000000}%
\pgfsetstrokecolor{textcolor}%
\pgfsetfillcolor{textcolor}%
\pgftext[x=5.660196in,y=0.640058in,,top]{\color{textcolor}\sffamily\fontsize{10.000000}{12.000000}\selectfont 0.12}%
\end{pgfscope}%
\begin{pgfscope}%
\pgfpathrectangle{\pgfqpoint{0.777361in}{0.737281in}}{\pgfqpoint{7.072639in}{2.612719in}}%
\pgfusepath{clip}%
\pgfsetrectcap%
\pgfsetroundjoin%
\pgfsetlinewidth{0.803000pt}%
\definecolor{currentstroke}{rgb}{0.690196,0.690196,0.690196}%
\pgfsetstrokecolor{currentstroke}%
\pgfsetstrokeopacity{0.700000}%
\pgfsetdash{}{0pt}%
\pgfpathmoveto{\pgfqpoint{6.512949in}{0.737281in}}%
\pgfpathlineto{\pgfqpoint{6.512949in}{3.350000in}}%
\pgfusepath{stroke}%
\end{pgfscope}%
\begin{pgfscope}%
\pgfsetbuttcap%
\pgfsetroundjoin%
\definecolor{currentfill}{rgb}{0.000000,0.000000,0.000000}%
\pgfsetfillcolor{currentfill}%
\pgfsetlinewidth{0.803000pt}%
\definecolor{currentstroke}{rgb}{0.000000,0.000000,0.000000}%
\pgfsetstrokecolor{currentstroke}%
\pgfsetdash{}{0pt}%
\pgfsys@defobject{currentmarker}{\pgfqpoint{0.000000in}{-0.048611in}}{\pgfqpoint{0.000000in}{0.000000in}}{%
\pgfpathmoveto{\pgfqpoint{0.000000in}{0.000000in}}%
\pgfpathlineto{\pgfqpoint{0.000000in}{-0.048611in}}%
\pgfusepath{stroke,fill}%
}%
\begin{pgfscope}%
\pgfsys@transformshift{6.512949in}{0.737281in}%
\pgfsys@useobject{currentmarker}{}%
\end{pgfscope}%
\end{pgfscope}%
\begin{pgfscope}%
\definecolor{textcolor}{rgb}{0.000000,0.000000,0.000000}%
\pgfsetstrokecolor{textcolor}%
\pgfsetfillcolor{textcolor}%
\pgftext[x=6.512949in,y=0.640058in,,top]{\color{textcolor}\sffamily\fontsize{10.000000}{12.000000}\selectfont 0.14}%
\end{pgfscope}%
\begin{pgfscope}%
\pgfpathrectangle{\pgfqpoint{0.777361in}{0.737281in}}{\pgfqpoint{7.072639in}{2.612719in}}%
\pgfusepath{clip}%
\pgfsetrectcap%
\pgfsetroundjoin%
\pgfsetlinewidth{0.803000pt}%
\definecolor{currentstroke}{rgb}{0.690196,0.690196,0.690196}%
\pgfsetstrokecolor{currentstroke}%
\pgfsetstrokeopacity{0.700000}%
\pgfsetdash{}{0pt}%
\pgfpathmoveto{\pgfqpoint{7.365702in}{0.737281in}}%
\pgfpathlineto{\pgfqpoint{7.365702in}{3.350000in}}%
\pgfusepath{stroke}%
\end{pgfscope}%
\begin{pgfscope}%
\pgfsetbuttcap%
\pgfsetroundjoin%
\definecolor{currentfill}{rgb}{0.000000,0.000000,0.000000}%
\pgfsetfillcolor{currentfill}%
\pgfsetlinewidth{0.803000pt}%
\definecolor{currentstroke}{rgb}{0.000000,0.000000,0.000000}%
\pgfsetstrokecolor{currentstroke}%
\pgfsetdash{}{0pt}%
\pgfsys@defobject{currentmarker}{\pgfqpoint{0.000000in}{-0.048611in}}{\pgfqpoint{0.000000in}{0.000000in}}{%
\pgfpathmoveto{\pgfqpoint{0.000000in}{0.000000in}}%
\pgfpathlineto{\pgfqpoint{0.000000in}{-0.048611in}}%
\pgfusepath{stroke,fill}%
}%
\begin{pgfscope}%
\pgfsys@transformshift{7.365702in}{0.737281in}%
\pgfsys@useobject{currentmarker}{}%
\end{pgfscope}%
\end{pgfscope}%
\begin{pgfscope}%
\definecolor{textcolor}{rgb}{0.000000,0.000000,0.000000}%
\pgfsetstrokecolor{textcolor}%
\pgfsetfillcolor{textcolor}%
\pgftext[x=7.365702in,y=0.640058in,,top]{\color{textcolor}\sffamily\fontsize{10.000000}{12.000000}\selectfont 0.16}%
\end{pgfscope}%
\begin{pgfscope}%
\definecolor{textcolor}{rgb}{0.000000,0.000000,0.000000}%
\pgfsetstrokecolor{textcolor}%
\pgfsetfillcolor{textcolor}%
\pgftext[x=4.313681in,y=0.450090in,,top]{\color{textcolor}\sffamily\fontsize{10.000000}{12.000000}\selectfont \(\displaystyle \frac{1}{2\kappa_a}-\frac{1}{2\kappa_a^{(c)}}\)}%
\end{pgfscope}%
\begin{pgfscope}%
\pgfpathrectangle{\pgfqpoint{0.777361in}{0.737281in}}{\pgfqpoint{7.072639in}{2.612719in}}%
\pgfusepath{clip}%
\pgfsetrectcap%
\pgfsetroundjoin%
\pgfsetlinewidth{0.803000pt}%
\definecolor{currentstroke}{rgb}{0.690196,0.690196,0.690196}%
\pgfsetstrokecolor{currentstroke}%
\pgfsetstrokeopacity{0.700000}%
\pgfsetdash{}{0pt}%
\pgfpathmoveto{\pgfqpoint{0.777361in}{0.838683in}}%
\pgfpathlineto{\pgfqpoint{7.850000in}{0.838683in}}%
\pgfusepath{stroke}%
\end{pgfscope}%
\begin{pgfscope}%
\pgfsetbuttcap%
\pgfsetroundjoin%
\definecolor{currentfill}{rgb}{0.000000,0.000000,0.000000}%
\pgfsetfillcolor{currentfill}%
\pgfsetlinewidth{0.803000pt}%
\definecolor{currentstroke}{rgb}{0.000000,0.000000,0.000000}%
\pgfsetstrokecolor{currentstroke}%
\pgfsetdash{}{0pt}%
\pgfsys@defobject{currentmarker}{\pgfqpoint{-0.048611in}{0.000000in}}{\pgfqpoint{-0.000000in}{0.000000in}}{%
\pgfpathmoveto{\pgfqpoint{-0.000000in}{0.000000in}}%
\pgfpathlineto{\pgfqpoint{-0.048611in}{0.000000in}}%
\pgfusepath{stroke,fill}%
}%
\begin{pgfscope}%
\pgfsys@transformshift{0.777361in}{0.838683in}%
\pgfsys@useobject{currentmarker}{}%
\end{pgfscope}%
\end{pgfscope}%
\begin{pgfscope}%
\definecolor{textcolor}{rgb}{0.000000,0.000000,0.000000}%
\pgfsetstrokecolor{textcolor}%
\pgfsetfillcolor{textcolor}%
\pgftext[x=0.459259in, y=0.785922in, left, base]{\color{textcolor}\sffamily\fontsize{10.000000}{12.000000}\selectfont 1.0}%
\end{pgfscope}%
\begin{pgfscope}%
\pgfpathrectangle{\pgfqpoint{0.777361in}{0.737281in}}{\pgfqpoint{7.072639in}{2.612719in}}%
\pgfusepath{clip}%
\pgfsetrectcap%
\pgfsetroundjoin%
\pgfsetlinewidth{0.803000pt}%
\definecolor{currentstroke}{rgb}{0.690196,0.690196,0.690196}%
\pgfsetstrokecolor{currentstroke}%
\pgfsetstrokeopacity{0.700000}%
\pgfsetdash{}{0pt}%
\pgfpathmoveto{\pgfqpoint{0.777361in}{1.398187in}}%
\pgfpathlineto{\pgfqpoint{7.850000in}{1.398187in}}%
\pgfusepath{stroke}%
\end{pgfscope}%
\begin{pgfscope}%
\pgfsetbuttcap%
\pgfsetroundjoin%
\definecolor{currentfill}{rgb}{0.000000,0.000000,0.000000}%
\pgfsetfillcolor{currentfill}%
\pgfsetlinewidth{0.803000pt}%
\definecolor{currentstroke}{rgb}{0.000000,0.000000,0.000000}%
\pgfsetstrokecolor{currentstroke}%
\pgfsetdash{}{0pt}%
\pgfsys@defobject{currentmarker}{\pgfqpoint{-0.048611in}{0.000000in}}{\pgfqpoint{-0.000000in}{0.000000in}}{%
\pgfpathmoveto{\pgfqpoint{-0.000000in}{0.000000in}}%
\pgfpathlineto{\pgfqpoint{-0.048611in}{0.000000in}}%
\pgfusepath{stroke,fill}%
}%
\begin{pgfscope}%
\pgfsys@transformshift{0.777361in}{1.398187in}%
\pgfsys@useobject{currentmarker}{}%
\end{pgfscope}%
\end{pgfscope}%
\begin{pgfscope}%
\definecolor{textcolor}{rgb}{0.000000,0.000000,0.000000}%
\pgfsetstrokecolor{textcolor}%
\pgfsetfillcolor{textcolor}%
\pgftext[x=0.459259in, y=1.345425in, left, base]{\color{textcolor}\sffamily\fontsize{10.000000}{12.000000}\selectfont 1.2}%
\end{pgfscope}%
\begin{pgfscope}%
\pgfpathrectangle{\pgfqpoint{0.777361in}{0.737281in}}{\pgfqpoint{7.072639in}{2.612719in}}%
\pgfusepath{clip}%
\pgfsetrectcap%
\pgfsetroundjoin%
\pgfsetlinewidth{0.803000pt}%
\definecolor{currentstroke}{rgb}{0.690196,0.690196,0.690196}%
\pgfsetstrokecolor{currentstroke}%
\pgfsetstrokeopacity{0.700000}%
\pgfsetdash{}{0pt}%
\pgfpathmoveto{\pgfqpoint{0.777361in}{1.957691in}}%
\pgfpathlineto{\pgfqpoint{7.850000in}{1.957691in}}%
\pgfusepath{stroke}%
\end{pgfscope}%
\begin{pgfscope}%
\pgfsetbuttcap%
\pgfsetroundjoin%
\definecolor{currentfill}{rgb}{0.000000,0.000000,0.000000}%
\pgfsetfillcolor{currentfill}%
\pgfsetlinewidth{0.803000pt}%
\definecolor{currentstroke}{rgb}{0.000000,0.000000,0.000000}%
\pgfsetstrokecolor{currentstroke}%
\pgfsetdash{}{0pt}%
\pgfsys@defobject{currentmarker}{\pgfqpoint{-0.048611in}{0.000000in}}{\pgfqpoint{-0.000000in}{0.000000in}}{%
\pgfpathmoveto{\pgfqpoint{-0.000000in}{0.000000in}}%
\pgfpathlineto{\pgfqpoint{-0.048611in}{0.000000in}}%
\pgfusepath{stroke,fill}%
}%
\begin{pgfscope}%
\pgfsys@transformshift{0.777361in}{1.957691in}%
\pgfsys@useobject{currentmarker}{}%
\end{pgfscope}%
\end{pgfscope}%
\begin{pgfscope}%
\definecolor{textcolor}{rgb}{0.000000,0.000000,0.000000}%
\pgfsetstrokecolor{textcolor}%
\pgfsetfillcolor{textcolor}%
\pgftext[x=0.459259in, y=1.904929in, left, base]{\color{textcolor}\sffamily\fontsize{10.000000}{12.000000}\selectfont 1.4}%
\end{pgfscope}%
\begin{pgfscope}%
\pgfpathrectangle{\pgfqpoint{0.777361in}{0.737281in}}{\pgfqpoint{7.072639in}{2.612719in}}%
\pgfusepath{clip}%
\pgfsetrectcap%
\pgfsetroundjoin%
\pgfsetlinewidth{0.803000pt}%
\definecolor{currentstroke}{rgb}{0.690196,0.690196,0.690196}%
\pgfsetstrokecolor{currentstroke}%
\pgfsetstrokeopacity{0.700000}%
\pgfsetdash{}{0pt}%
\pgfpathmoveto{\pgfqpoint{0.777361in}{2.517195in}}%
\pgfpathlineto{\pgfqpoint{7.850000in}{2.517195in}}%
\pgfusepath{stroke}%
\end{pgfscope}%
\begin{pgfscope}%
\pgfsetbuttcap%
\pgfsetroundjoin%
\definecolor{currentfill}{rgb}{0.000000,0.000000,0.000000}%
\pgfsetfillcolor{currentfill}%
\pgfsetlinewidth{0.803000pt}%
\definecolor{currentstroke}{rgb}{0.000000,0.000000,0.000000}%
\pgfsetstrokecolor{currentstroke}%
\pgfsetdash{}{0pt}%
\pgfsys@defobject{currentmarker}{\pgfqpoint{-0.048611in}{0.000000in}}{\pgfqpoint{-0.000000in}{0.000000in}}{%
\pgfpathmoveto{\pgfqpoint{-0.000000in}{0.000000in}}%
\pgfpathlineto{\pgfqpoint{-0.048611in}{0.000000in}}%
\pgfusepath{stroke,fill}%
}%
\begin{pgfscope}%
\pgfsys@transformshift{0.777361in}{2.517195in}%
\pgfsys@useobject{currentmarker}{}%
\end{pgfscope}%
\end{pgfscope}%
\begin{pgfscope}%
\definecolor{textcolor}{rgb}{0.000000,0.000000,0.000000}%
\pgfsetstrokecolor{textcolor}%
\pgfsetfillcolor{textcolor}%
\pgftext[x=0.459259in, y=2.464433in, left, base]{\color{textcolor}\sffamily\fontsize{10.000000}{12.000000}\selectfont 1.6}%
\end{pgfscope}%
\begin{pgfscope}%
\pgfpathrectangle{\pgfqpoint{0.777361in}{0.737281in}}{\pgfqpoint{7.072639in}{2.612719in}}%
\pgfusepath{clip}%
\pgfsetrectcap%
\pgfsetroundjoin%
\pgfsetlinewidth{0.803000pt}%
\definecolor{currentstroke}{rgb}{0.690196,0.690196,0.690196}%
\pgfsetstrokecolor{currentstroke}%
\pgfsetstrokeopacity{0.700000}%
\pgfsetdash{}{0pt}%
\pgfpathmoveto{\pgfqpoint{0.777361in}{3.076698in}}%
\pgfpathlineto{\pgfqpoint{7.850000in}{3.076698in}}%
\pgfusepath{stroke}%
\end{pgfscope}%
\begin{pgfscope}%
\pgfsetbuttcap%
\pgfsetroundjoin%
\definecolor{currentfill}{rgb}{0.000000,0.000000,0.000000}%
\pgfsetfillcolor{currentfill}%
\pgfsetlinewidth{0.803000pt}%
\definecolor{currentstroke}{rgb}{0.000000,0.000000,0.000000}%
\pgfsetstrokecolor{currentstroke}%
\pgfsetdash{}{0pt}%
\pgfsys@defobject{currentmarker}{\pgfqpoint{-0.048611in}{0.000000in}}{\pgfqpoint{-0.000000in}{0.000000in}}{%
\pgfpathmoveto{\pgfqpoint{-0.000000in}{0.000000in}}%
\pgfpathlineto{\pgfqpoint{-0.048611in}{0.000000in}}%
\pgfusepath{stroke,fill}%
}%
\begin{pgfscope}%
\pgfsys@transformshift{0.777361in}{3.076698in}%
\pgfsys@useobject{currentmarker}{}%
\end{pgfscope}%
\end{pgfscope}%
\begin{pgfscope}%
\definecolor{textcolor}{rgb}{0.000000,0.000000,0.000000}%
\pgfsetstrokecolor{textcolor}%
\pgfsetfillcolor{textcolor}%
\pgftext[x=0.459259in, y=3.023937in, left, base]{\color{textcolor}\sffamily\fontsize{10.000000}{12.000000}\selectfont 1.8}%
\end{pgfscope}%
\begin{pgfscope}%
\definecolor{textcolor}{rgb}{0.000000,0.000000,0.000000}%
\pgfsetstrokecolor{textcolor}%
\pgfsetfillcolor{textcolor}%
\pgftext[x=0.403704in,y=2.043640in,,bottom,rotate=90.000000]{\color{textcolor}\sffamily\fontsize{10.000000}{12.000000}\selectfont \(\displaystyle \sqrt{8t_1}m_\rho^\chi\)}%
\end{pgfscope}%
\begin{pgfscope}%
\pgfpathrectangle{\pgfqpoint{0.777361in}{0.737281in}}{\pgfqpoint{7.072639in}{2.612719in}}%
\pgfusepath{clip}%
\pgfsetbuttcap%
\pgfsetroundjoin%
\pgfsetlinewidth{1.505625pt}%
\definecolor{currentstroke}{rgb}{0.121569,0.466667,0.705882}%
\pgfsetstrokecolor{currentstroke}%
\pgfsetdash{}{0pt}%
\pgfpathmoveto{\pgfqpoint{3.218112in}{2.562596in}}%
\pgfpathlineto{\pgfqpoint{3.263602in}{2.562596in}}%
\pgfusepath{stroke}%
\end{pgfscope}%
\begin{pgfscope}%
\pgfpathrectangle{\pgfqpoint{0.777361in}{0.737281in}}{\pgfqpoint{7.072639in}{2.612719in}}%
\pgfusepath{clip}%
\pgfsetbuttcap%
\pgfsetroundjoin%
\pgfsetlinewidth{1.505625pt}%
\definecolor{currentstroke}{rgb}{0.121569,0.466667,0.705882}%
\pgfsetstrokecolor{currentstroke}%
\pgfsetdash{}{0pt}%
\pgfpathmoveto{\pgfqpoint{2.036981in}{1.729855in}}%
\pgfpathlineto{\pgfqpoint{2.082470in}{1.729855in}}%
\pgfusepath{stroke}%
\end{pgfscope}%
\begin{pgfscope}%
\pgfpathrectangle{\pgfqpoint{0.777361in}{0.737281in}}{\pgfqpoint{7.072639in}{2.612719in}}%
\pgfusepath{clip}%
\pgfsetbuttcap%
\pgfsetroundjoin%
\pgfsetlinewidth{1.505625pt}%
\definecolor{currentstroke}{rgb}{0.121569,0.466667,0.705882}%
\pgfsetstrokecolor{currentstroke}%
\pgfsetdash{}{0pt}%
\pgfpathmoveto{\pgfqpoint{3.240857in}{1.893952in}}%
\pgfpathlineto{\pgfqpoint{3.240857in}{3.231240in}}%
\pgfusepath{stroke}%
\end{pgfscope}%
\begin{pgfscope}%
\pgfpathrectangle{\pgfqpoint{0.777361in}{0.737281in}}{\pgfqpoint{7.072639in}{2.612719in}}%
\pgfusepath{clip}%
\pgfsetbuttcap%
\pgfsetroundjoin%
\pgfsetlinewidth{1.505625pt}%
\definecolor{currentstroke}{rgb}{0.121569,0.466667,0.705882}%
\pgfsetstrokecolor{currentstroke}%
\pgfsetdash{}{0pt}%
\pgfpathmoveto{\pgfqpoint{2.059726in}{1.280092in}}%
\pgfpathlineto{\pgfqpoint{2.059726in}{2.179617in}}%
\pgfusepath{stroke}%
\end{pgfscope}%
\begin{pgfscope}%
\pgfpathrectangle{\pgfqpoint{0.777361in}{0.737281in}}{\pgfqpoint{7.072639in}{2.612719in}}%
\pgfusepath{clip}%
\pgfsetbuttcap%
\pgfsetroundjoin%
\pgfsetlinewidth{1.505625pt}%
\definecolor{currentstroke}{rgb}{1.000000,0.498039,0.054902}%
\pgfsetstrokecolor{currentstroke}%
\pgfsetdash{}{0pt}%
\pgfpathmoveto{\pgfqpoint{7.504502in}{2.242291in}}%
\pgfpathlineto{\pgfqpoint{7.528516in}{2.242291in}}%
\pgfusepath{stroke}%
\end{pgfscope}%
\begin{pgfscope}%
\pgfpathrectangle{\pgfqpoint{0.777361in}{0.737281in}}{\pgfqpoint{7.072639in}{2.612719in}}%
\pgfusepath{clip}%
\pgfsetbuttcap%
\pgfsetroundjoin%
\pgfsetlinewidth{1.505625pt}%
\definecolor{currentstroke}{rgb}{1.000000,0.498039,0.054902}%
\pgfsetstrokecolor{currentstroke}%
\pgfsetdash{}{0pt}%
\pgfpathmoveto{\pgfqpoint{5.836362in}{2.116547in}}%
\pgfpathlineto{\pgfqpoint{5.860377in}{2.116547in}}%
\pgfusepath{stroke}%
\end{pgfscope}%
\begin{pgfscope}%
\pgfpathrectangle{\pgfqpoint{0.777361in}{0.737281in}}{\pgfqpoint{7.072639in}{2.612719in}}%
\pgfusepath{clip}%
\pgfsetbuttcap%
\pgfsetroundjoin%
\pgfsetlinewidth{1.505625pt}%
\definecolor{currentstroke}{rgb}{1.000000,0.498039,0.054902}%
\pgfsetstrokecolor{currentstroke}%
\pgfsetdash{}{0pt}%
\pgfpathmoveto{\pgfqpoint{4.213924in}{1.674364in}}%
\pgfpathlineto{\pgfqpoint{4.237939in}{1.674364in}}%
\pgfusepath{stroke}%
\end{pgfscope}%
\begin{pgfscope}%
\pgfpathrectangle{\pgfqpoint{0.777361in}{0.737281in}}{\pgfqpoint{7.072639in}{2.612719in}}%
\pgfusepath{clip}%
\pgfsetbuttcap%
\pgfsetroundjoin%
\pgfsetlinewidth{1.505625pt}%
\definecolor{currentstroke}{rgb}{1.000000,0.498039,0.054902}%
\pgfsetstrokecolor{currentstroke}%
\pgfsetdash{}{0pt}%
\pgfpathmoveto{\pgfqpoint{2.635337in}{1.247180in}}%
\pgfpathlineto{\pgfqpoint{2.659351in}{1.247180in}}%
\pgfusepath{stroke}%
\end{pgfscope}%
\begin{pgfscope}%
\pgfpathrectangle{\pgfqpoint{0.777361in}{0.737281in}}{\pgfqpoint{7.072639in}{2.612719in}}%
\pgfusepath{clip}%
\pgfsetbuttcap%
\pgfsetroundjoin%
\pgfsetlinewidth{1.505625pt}%
\definecolor{currentstroke}{rgb}{1.000000,0.498039,0.054902}%
\pgfsetstrokecolor{currentstroke}%
\pgfsetdash{}{0pt}%
\pgfpathmoveto{\pgfqpoint{1.098845in}{1.665805in}}%
\pgfpathlineto{\pgfqpoint{1.122859in}{1.665805in}}%
\pgfusepath{stroke}%
\end{pgfscope}%
\begin{pgfscope}%
\pgfpathrectangle{\pgfqpoint{0.777361in}{0.737281in}}{\pgfqpoint{7.072639in}{2.612719in}}%
\pgfusepath{clip}%
\pgfsetbuttcap%
\pgfsetroundjoin%
\pgfsetlinewidth{1.505625pt}%
\definecolor{currentstroke}{rgb}{1.000000,0.498039,0.054902}%
\pgfsetstrokecolor{currentstroke}%
\pgfsetdash{}{0pt}%
\pgfpathmoveto{\pgfqpoint{7.516509in}{1.969619in}}%
\pgfpathlineto{\pgfqpoint{7.516509in}{2.514963in}}%
\pgfusepath{stroke}%
\end{pgfscope}%
\begin{pgfscope}%
\pgfpathrectangle{\pgfqpoint{0.777361in}{0.737281in}}{\pgfqpoint{7.072639in}{2.612719in}}%
\pgfusepath{clip}%
\pgfsetbuttcap%
\pgfsetroundjoin%
\pgfsetlinewidth{1.505625pt}%
\definecolor{currentstroke}{rgb}{1.000000,0.498039,0.054902}%
\pgfsetstrokecolor{currentstroke}%
\pgfsetdash{}{0pt}%
\pgfpathmoveto{\pgfqpoint{5.848369in}{1.763590in}}%
\pgfpathlineto{\pgfqpoint{5.848369in}{2.469504in}}%
\pgfusepath{stroke}%
\end{pgfscope}%
\begin{pgfscope}%
\pgfpathrectangle{\pgfqpoint{0.777361in}{0.737281in}}{\pgfqpoint{7.072639in}{2.612719in}}%
\pgfusepath{clip}%
\pgfsetbuttcap%
\pgfsetroundjoin%
\pgfsetlinewidth{1.505625pt}%
\definecolor{currentstroke}{rgb}{1.000000,0.498039,0.054902}%
\pgfsetstrokecolor{currentstroke}%
\pgfsetdash{}{0pt}%
\pgfpathmoveto{\pgfqpoint{4.225932in}{1.349372in}}%
\pgfpathlineto{\pgfqpoint{4.225932in}{1.999357in}}%
\pgfusepath{stroke}%
\end{pgfscope}%
\begin{pgfscope}%
\pgfpathrectangle{\pgfqpoint{0.777361in}{0.737281in}}{\pgfqpoint{7.072639in}{2.612719in}}%
\pgfusepath{clip}%
\pgfsetbuttcap%
\pgfsetroundjoin%
\pgfsetlinewidth{1.505625pt}%
\definecolor{currentstroke}{rgb}{1.000000,0.498039,0.054902}%
\pgfsetstrokecolor{currentstroke}%
\pgfsetdash{}{0pt}%
\pgfpathmoveto{\pgfqpoint{2.647344in}{0.856041in}}%
\pgfpathlineto{\pgfqpoint{2.647344in}{1.638319in}}%
\pgfusepath{stroke}%
\end{pgfscope}%
\begin{pgfscope}%
\pgfpathrectangle{\pgfqpoint{0.777361in}{0.737281in}}{\pgfqpoint{7.072639in}{2.612719in}}%
\pgfusepath{clip}%
\pgfsetbuttcap%
\pgfsetroundjoin%
\pgfsetlinewidth{1.505625pt}%
\definecolor{currentstroke}{rgb}{1.000000,0.498039,0.054902}%
\pgfsetstrokecolor{currentstroke}%
\pgfsetdash{}{0pt}%
\pgfpathmoveto{\pgfqpoint{1.110852in}{1.239341in}}%
\pgfpathlineto{\pgfqpoint{1.110852in}{2.092269in}}%
\pgfusepath{stroke}%
\end{pgfscope}%
\begin{pgfscope}%
\pgfpathrectangle{\pgfqpoint{0.777361in}{0.737281in}}{\pgfqpoint{7.072639in}{2.612719in}}%
\pgfusepath{clip}%
\pgfsetrectcap%
\pgfsetroundjoin%
\pgfsetlinewidth{1.505625pt}%
\definecolor{currentstroke}{rgb}{1.000000,0.000000,0.000000}%
\pgfsetstrokecolor{currentstroke}%
\pgfsetstrokeopacity{0.500000}%
\pgfsetdash{}{0pt}%
\pgfpathmoveto{\pgfqpoint{0.777361in}{1.885882in}}%
\pgfpathlineto{\pgfqpoint{7.850000in}{1.885882in}}%
\pgfusepath{stroke}%
\end{pgfscope}%
\begin{pgfscope}%
\pgfpathrectangle{\pgfqpoint{0.777361in}{0.737281in}}{\pgfqpoint{7.072639in}{2.612719in}}%
\pgfusepath{clip}%
\pgfsetbuttcap%
\pgfsetroundjoin%
\definecolor{currentfill}{rgb}{0.121569,0.466667,0.705882}%
\pgfsetfillcolor{currentfill}%
\pgfsetlinewidth{1.003750pt}%
\definecolor{currentstroke}{rgb}{0.121569,0.466667,0.705882}%
\pgfsetstrokecolor{currentstroke}%
\pgfsetdash{}{0pt}%
\pgfsys@defobject{currentmarker}{\pgfqpoint{-0.020833in}{-0.020833in}}{\pgfqpoint{0.020833in}{0.020833in}}{%
\pgfpathmoveto{\pgfqpoint{0.000000in}{-0.020833in}}%
\pgfpathcurveto{\pgfqpoint{0.005525in}{-0.020833in}}{\pgfqpoint{0.010825in}{-0.018638in}}{\pgfqpoint{0.014731in}{-0.014731in}}%
\pgfpathcurveto{\pgfqpoint{0.018638in}{-0.010825in}}{\pgfqpoint{0.020833in}{-0.005525in}}{\pgfqpoint{0.020833in}{0.000000in}}%
\pgfpathcurveto{\pgfqpoint{0.020833in}{0.005525in}}{\pgfqpoint{0.018638in}{0.010825in}}{\pgfqpoint{0.014731in}{0.014731in}}%
\pgfpathcurveto{\pgfqpoint{0.010825in}{0.018638in}}{\pgfqpoint{0.005525in}{0.020833in}}{\pgfqpoint{0.000000in}{0.020833in}}%
\pgfpathcurveto{\pgfqpoint{-0.005525in}{0.020833in}}{\pgfqpoint{-0.010825in}{0.018638in}}{\pgfqpoint{-0.014731in}{0.014731in}}%
\pgfpathcurveto{\pgfqpoint{-0.018638in}{0.010825in}}{\pgfqpoint{-0.020833in}{0.005525in}}{\pgfqpoint{-0.020833in}{0.000000in}}%
\pgfpathcurveto{\pgfqpoint{-0.020833in}{-0.005525in}}{\pgfqpoint{-0.018638in}{-0.010825in}}{\pgfqpoint{-0.014731in}{-0.014731in}}%
\pgfpathcurveto{\pgfqpoint{-0.010825in}{-0.018638in}}{\pgfqpoint{-0.005525in}{-0.020833in}}{\pgfqpoint{0.000000in}{-0.020833in}}%
\pgfpathclose%
\pgfusepath{stroke,fill}%
}%
\begin{pgfscope}%
\pgfsys@transformshift{3.240857in}{2.562596in}%
\pgfsys@useobject{currentmarker}{}%
\end{pgfscope}%
\begin{pgfscope}%
\pgfsys@transformshift{2.059726in}{1.729855in}%
\pgfsys@useobject{currentmarker}{}%
\end{pgfscope}%
\end{pgfscope}%
\begin{pgfscope}%
\pgfpathrectangle{\pgfqpoint{0.777361in}{0.737281in}}{\pgfqpoint{7.072639in}{2.612719in}}%
\pgfusepath{clip}%
\pgfsetbuttcap%
\pgfsetroundjoin%
\definecolor{currentfill}{rgb}{1.000000,0.498039,0.054902}%
\pgfsetfillcolor{currentfill}%
\pgfsetlinewidth{1.003750pt}%
\definecolor{currentstroke}{rgb}{1.000000,0.498039,0.054902}%
\pgfsetstrokecolor{currentstroke}%
\pgfsetdash{}{0pt}%
\pgfsys@defobject{currentmarker}{\pgfqpoint{-0.020833in}{-0.020833in}}{\pgfqpoint{0.020833in}{0.020833in}}{%
\pgfpathmoveto{\pgfqpoint{0.000000in}{-0.020833in}}%
\pgfpathcurveto{\pgfqpoint{0.005525in}{-0.020833in}}{\pgfqpoint{0.010825in}{-0.018638in}}{\pgfqpoint{0.014731in}{-0.014731in}}%
\pgfpathcurveto{\pgfqpoint{0.018638in}{-0.010825in}}{\pgfqpoint{0.020833in}{-0.005525in}}{\pgfqpoint{0.020833in}{0.000000in}}%
\pgfpathcurveto{\pgfqpoint{0.020833in}{0.005525in}}{\pgfqpoint{0.018638in}{0.010825in}}{\pgfqpoint{0.014731in}{0.014731in}}%
\pgfpathcurveto{\pgfqpoint{0.010825in}{0.018638in}}{\pgfqpoint{0.005525in}{0.020833in}}{\pgfqpoint{0.000000in}{0.020833in}}%
\pgfpathcurveto{\pgfqpoint{-0.005525in}{0.020833in}}{\pgfqpoint{-0.010825in}{0.018638in}}{\pgfqpoint{-0.014731in}{0.014731in}}%
\pgfpathcurveto{\pgfqpoint{-0.018638in}{0.010825in}}{\pgfqpoint{-0.020833in}{0.005525in}}{\pgfqpoint{-0.020833in}{0.000000in}}%
\pgfpathcurveto{\pgfqpoint{-0.020833in}{-0.005525in}}{\pgfqpoint{-0.018638in}{-0.010825in}}{\pgfqpoint{-0.014731in}{-0.014731in}}%
\pgfpathcurveto{\pgfqpoint{-0.010825in}{-0.018638in}}{\pgfqpoint{-0.005525in}{-0.020833in}}{\pgfqpoint{0.000000in}{-0.020833in}}%
\pgfpathclose%
\pgfusepath{stroke,fill}%
}%
\begin{pgfscope}%
\pgfsys@transformshift{7.516509in}{2.242291in}%
\pgfsys@useobject{currentmarker}{}%
\end{pgfscope}%
\begin{pgfscope}%
\pgfsys@transformshift{5.848369in}{2.116547in}%
\pgfsys@useobject{currentmarker}{}%
\end{pgfscope}%
\begin{pgfscope}%
\pgfsys@transformshift{4.225932in}{1.674364in}%
\pgfsys@useobject{currentmarker}{}%
\end{pgfscope}%
\begin{pgfscope}%
\pgfsys@transformshift{2.647344in}{1.247180in}%
\pgfsys@useobject{currentmarker}{}%
\end{pgfscope}%
\begin{pgfscope}%
\pgfsys@transformshift{1.110852in}{1.665805in}%
\pgfsys@useobject{currentmarker}{}%
\end{pgfscope}%
\end{pgfscope}%
\begin{pgfscope}%
\pgfsetrectcap%
\pgfsetmiterjoin%
\pgfsetlinewidth{0.803000pt}%
\definecolor{currentstroke}{rgb}{0.000000,0.000000,0.000000}%
\pgfsetstrokecolor{currentstroke}%
\pgfsetdash{}{0pt}%
\pgfpathmoveto{\pgfqpoint{0.777361in}{0.737281in}}%
\pgfpathlineto{\pgfqpoint{0.777361in}{3.350000in}}%
\pgfusepath{stroke}%
\end{pgfscope}%
\begin{pgfscope}%
\pgfsetrectcap%
\pgfsetmiterjoin%
\pgfsetlinewidth{0.803000pt}%
\definecolor{currentstroke}{rgb}{0.000000,0.000000,0.000000}%
\pgfsetstrokecolor{currentstroke}%
\pgfsetdash{}{0pt}%
\pgfpathmoveto{\pgfqpoint{7.850000in}{0.737281in}}%
\pgfpathlineto{\pgfqpoint{7.850000in}{3.350000in}}%
\pgfusepath{stroke}%
\end{pgfscope}%
\begin{pgfscope}%
\pgfsetrectcap%
\pgfsetmiterjoin%
\pgfsetlinewidth{0.803000pt}%
\definecolor{currentstroke}{rgb}{0.000000,0.000000,0.000000}%
\pgfsetstrokecolor{currentstroke}%
\pgfsetdash{}{0pt}%
\pgfpathmoveto{\pgfqpoint{0.777361in}{0.737281in}}%
\pgfpathlineto{\pgfqpoint{7.850000in}{0.737281in}}%
\pgfusepath{stroke}%
\end{pgfscope}%
\begin{pgfscope}%
\pgfsetrectcap%
\pgfsetmiterjoin%
\pgfsetlinewidth{0.803000pt}%
\definecolor{currentstroke}{rgb}{0.000000,0.000000,0.000000}%
\pgfsetstrokecolor{currentstroke}%
\pgfsetdash{}{0pt}%
\pgfpathmoveto{\pgfqpoint{0.777361in}{3.350000in}}%
\pgfpathlineto{\pgfqpoint{7.850000in}{3.350000in}}%
\pgfusepath{stroke}%
\end{pgfscope}%
\begin{pgfscope}%
\pgfsetbuttcap%
\pgfsetmiterjoin%
\definecolor{currentfill}{rgb}{1.000000,1.000000,1.000000}%
\pgfsetfillcolor{currentfill}%
\pgfsetfillopacity{0.800000}%
\pgfsetlinewidth{1.003750pt}%
\definecolor{currentstroke}{rgb}{0.800000,0.800000,0.800000}%
\pgfsetstrokecolor{currentstroke}%
\pgfsetstrokeopacity{0.800000}%
\pgfsetdash{}{0pt}%
\pgfpathmoveto{\pgfqpoint{0.874583in}{2.614723in}}%
\pgfpathlineto{\pgfqpoint{1.969779in}{2.614723in}}%
\pgfpathquadraticcurveto{\pgfqpoint{1.997557in}{2.614723in}}{\pgfqpoint{1.997557in}{2.642501in}}%
\pgfpathlineto{\pgfqpoint{1.997557in}{3.252778in}}%
\pgfpathquadraticcurveto{\pgfqpoint{1.997557in}{3.280556in}}{\pgfqpoint{1.969779in}{3.280556in}}%
\pgfpathlineto{\pgfqpoint{0.874583in}{3.280556in}}%
\pgfpathquadraticcurveto{\pgfqpoint{0.846806in}{3.280556in}}{\pgfqpoint{0.846806in}{3.252778in}}%
\pgfpathlineto{\pgfqpoint{0.846806in}{2.642501in}}%
\pgfpathquadraticcurveto{\pgfqpoint{0.846806in}{2.614723in}}{\pgfqpoint{0.874583in}{2.614723in}}%
\pgfpathclose%
\pgfusepath{stroke,fill}%
\end{pgfscope}%
\begin{pgfscope}%
\pgfsetrectcap%
\pgfsetroundjoin%
\pgfsetlinewidth{1.505625pt}%
\definecolor{currentstroke}{rgb}{1.000000,0.000000,0.000000}%
\pgfsetstrokecolor{currentstroke}%
\pgfsetstrokeopacity{0.500000}%
\pgfsetdash{}{0pt}%
\pgfpathmoveto{\pgfqpoint{0.902361in}{3.155494in}}%
\pgfpathlineto{\pgfqpoint{1.180139in}{3.155494in}}%
\pgfusepath{stroke}%
\end{pgfscope}%
\begin{pgfscope}%
\definecolor{textcolor}{rgb}{0.000000,0.000000,0.000000}%
\pgfsetstrokecolor{textcolor}%
\pgfsetfillcolor{textcolor}%
\pgftext[x=1.291250in,y=3.106883in,left,base]{\color{textcolor}\sffamily\fontsize{10.000000}{12.000000}\selectfont \(\displaystyle \chi^2=1.106\)}%
\end{pgfscope}%
\begin{pgfscope}%
\pgfsetbuttcap%
\pgfsetroundjoin%
\pgfsetlinewidth{1.505625pt}%
\definecolor{currentstroke}{rgb}{0.121569,0.466667,0.705882}%
\pgfsetstrokecolor{currentstroke}%
\pgfsetdash{}{0pt}%
\pgfpathmoveto{\pgfqpoint{0.971806in}{2.951637in}}%
\pgfpathlineto{\pgfqpoint{1.110694in}{2.951637in}}%
\pgfusepath{stroke}%
\end{pgfscope}%
\begin{pgfscope}%
\pgfsetbuttcap%
\pgfsetroundjoin%
\pgfsetlinewidth{1.505625pt}%
\definecolor{currentstroke}{rgb}{0.121569,0.466667,0.705882}%
\pgfsetstrokecolor{currentstroke}%
\pgfsetdash{}{0pt}%
\pgfpathmoveto{\pgfqpoint{1.041250in}{2.882193in}}%
\pgfpathlineto{\pgfqpoint{1.041250in}{3.021082in}}%
\pgfusepath{stroke}%
\end{pgfscope}%
\begin{pgfscope}%
\pgfsetbuttcap%
\pgfsetroundjoin%
\definecolor{currentfill}{rgb}{0.121569,0.466667,0.705882}%
\pgfsetfillcolor{currentfill}%
\pgfsetlinewidth{1.003750pt}%
\definecolor{currentstroke}{rgb}{0.121569,0.466667,0.705882}%
\pgfsetstrokecolor{currentstroke}%
\pgfsetdash{}{0pt}%
\pgfsys@defobject{currentmarker}{\pgfqpoint{-0.020833in}{-0.020833in}}{\pgfqpoint{0.020833in}{0.020833in}}{%
\pgfpathmoveto{\pgfqpoint{0.000000in}{-0.020833in}}%
\pgfpathcurveto{\pgfqpoint{0.005525in}{-0.020833in}}{\pgfqpoint{0.010825in}{-0.018638in}}{\pgfqpoint{0.014731in}{-0.014731in}}%
\pgfpathcurveto{\pgfqpoint{0.018638in}{-0.010825in}}{\pgfqpoint{0.020833in}{-0.005525in}}{\pgfqpoint{0.020833in}{0.000000in}}%
\pgfpathcurveto{\pgfqpoint{0.020833in}{0.005525in}}{\pgfqpoint{0.018638in}{0.010825in}}{\pgfqpoint{0.014731in}{0.014731in}}%
\pgfpathcurveto{\pgfqpoint{0.010825in}{0.018638in}}{\pgfqpoint{0.005525in}{0.020833in}}{\pgfqpoint{0.000000in}{0.020833in}}%
\pgfpathcurveto{\pgfqpoint{-0.005525in}{0.020833in}}{\pgfqpoint{-0.010825in}{0.018638in}}{\pgfqpoint{-0.014731in}{0.014731in}}%
\pgfpathcurveto{\pgfqpoint{-0.018638in}{0.010825in}}{\pgfqpoint{-0.020833in}{0.005525in}}{\pgfqpoint{-0.020833in}{0.000000in}}%
\pgfpathcurveto{\pgfqpoint{-0.020833in}{-0.005525in}}{\pgfqpoint{-0.018638in}{-0.010825in}}{\pgfqpoint{-0.014731in}{-0.014731in}}%
\pgfpathcurveto{\pgfqpoint{-0.010825in}{-0.018638in}}{\pgfqpoint{-0.005525in}{-0.020833in}}{\pgfqpoint{0.000000in}{-0.020833in}}%
\pgfpathclose%
\pgfusepath{stroke,fill}%
}%
\begin{pgfscope}%
\pgfsys@transformshift{1.041250in}{2.951637in}%
\pgfsys@useobject{currentmarker}{}%
\end{pgfscope}%
\end{pgfscope}%
\begin{pgfscope}%
\definecolor{textcolor}{rgb}{0.000000,0.000000,0.000000}%
\pgfsetstrokecolor{textcolor}%
\pgfsetfillcolor{textcolor}%
\pgftext[x=1.291250in,y=2.903026in,left,base]{\color{textcolor}\sffamily\fontsize{10.000000}{12.000000}\selectfont \(\displaystyle b=0.34\)}%
\end{pgfscope}%
\begin{pgfscope}%
\pgfsetbuttcap%
\pgfsetroundjoin%
\pgfsetlinewidth{1.505625pt}%
\definecolor{currentstroke}{rgb}{1.000000,0.498039,0.054902}%
\pgfsetstrokecolor{currentstroke}%
\pgfsetdash{}{0pt}%
\pgfpathmoveto{\pgfqpoint{0.971806in}{2.747780in}}%
\pgfpathlineto{\pgfqpoint{1.110694in}{2.747780in}}%
\pgfusepath{stroke}%
\end{pgfscope}%
\begin{pgfscope}%
\pgfsetbuttcap%
\pgfsetroundjoin%
\pgfsetlinewidth{1.505625pt}%
\definecolor{currentstroke}{rgb}{1.000000,0.498039,0.054902}%
\pgfsetstrokecolor{currentstroke}%
\pgfsetdash{}{0pt}%
\pgfpathmoveto{\pgfqpoint{1.041250in}{2.678335in}}%
\pgfpathlineto{\pgfqpoint{1.041250in}{2.817224in}}%
\pgfusepath{stroke}%
\end{pgfscope}%
\begin{pgfscope}%
\pgfsetbuttcap%
\pgfsetroundjoin%
\definecolor{currentfill}{rgb}{1.000000,0.498039,0.054902}%
\pgfsetfillcolor{currentfill}%
\pgfsetlinewidth{1.003750pt}%
\definecolor{currentstroke}{rgb}{1.000000,0.498039,0.054902}%
\pgfsetstrokecolor{currentstroke}%
\pgfsetdash{}{0pt}%
\pgfsys@defobject{currentmarker}{\pgfqpoint{-0.020833in}{-0.020833in}}{\pgfqpoint{0.020833in}{0.020833in}}{%
\pgfpathmoveto{\pgfqpoint{0.000000in}{-0.020833in}}%
\pgfpathcurveto{\pgfqpoint{0.005525in}{-0.020833in}}{\pgfqpoint{0.010825in}{-0.018638in}}{\pgfqpoint{0.014731in}{-0.014731in}}%
\pgfpathcurveto{\pgfqpoint{0.018638in}{-0.010825in}}{\pgfqpoint{0.020833in}{-0.005525in}}{\pgfqpoint{0.020833in}{0.000000in}}%
\pgfpathcurveto{\pgfqpoint{0.020833in}{0.005525in}}{\pgfqpoint{0.018638in}{0.010825in}}{\pgfqpoint{0.014731in}{0.014731in}}%
\pgfpathcurveto{\pgfqpoint{0.010825in}{0.018638in}}{\pgfqpoint{0.005525in}{0.020833in}}{\pgfqpoint{0.000000in}{0.020833in}}%
\pgfpathcurveto{\pgfqpoint{-0.005525in}{0.020833in}}{\pgfqpoint{-0.010825in}{0.018638in}}{\pgfqpoint{-0.014731in}{0.014731in}}%
\pgfpathcurveto{\pgfqpoint{-0.018638in}{0.010825in}}{\pgfqpoint{-0.020833in}{0.005525in}}{\pgfqpoint{-0.020833in}{0.000000in}}%
\pgfpathcurveto{\pgfqpoint{-0.020833in}{-0.005525in}}{\pgfqpoint{-0.018638in}{-0.010825in}}{\pgfqpoint{-0.014731in}{-0.014731in}}%
\pgfpathcurveto{\pgfqpoint{-0.010825in}{-0.018638in}}{\pgfqpoint{-0.005525in}{-0.020833in}}{\pgfqpoint{0.000000in}{-0.020833in}}%
\pgfpathclose%
\pgfusepath{stroke,fill}%
}%
\begin{pgfscope}%
\pgfsys@transformshift{1.041250in}{2.747780in}%
\pgfsys@useobject{currentmarker}{}%
\end{pgfscope}%
\end{pgfscope}%
\begin{pgfscope}%
\definecolor{textcolor}{rgb}{0.000000,0.000000,0.000000}%
\pgfsetstrokecolor{textcolor}%
\pgfsetfillcolor{textcolor}%
\pgftext[x=1.291250in,y=2.699169in,left,base]{\color{textcolor}\sffamily\fontsize{10.000000}{12.000000}\selectfont \(\displaystyle b=0.35\)}%
\end{pgfscope}%
\end{pgfpicture}%
\makeatother%
\endgroup%

%% file: tek_conclusion.tex
\section{Conclusions}
We have performed the scale setting for $\mathcal{N}=1$ SUSY Yang-Mills at large-$N$ with Wilson flow methods. We were able to reduce the systematic errors related with finite-$N$ effects. The main results are summarized by Fig. \ref{fig:aw0}, in which we perform an extrapolation of the scale to the chiral limit. We showed that in the chiral limit, our results are in agreement with the prediction coming from the $\beta$-function of a $SU(N)$ Yang-Mills theory coupled with one Majorana fermion in the adjoint representation.

 As shown in Fig. \ref{fig:mrho_mpcac}, we also extracted the masses from the light-sector of the fundamental meson spectrum, and we verified that $\rho$-meson masses in the chiral limit are observables that can be used to set the scale of the theory. At this stage, the associated uncertainties do not allow one to use these extrapolated chiral masses as precision scales, nevertheless this is a nice crosscheck of the Wilson flow-related scales.